# A search for double-electron capture in $^{74}$Se using coincidence/anticoincidence gamma-ray spectrometry

M. Ješkovský[a], D. Frekers[b], A. Kováčik[a], P. P. Povinec[a*], P. Puppe[b], J. Staníček[a], I. Sýkora[a], F. Šimkovic[a,c], J.H. Thies[b]

[a] *Comenius University, Faculty of Mathematics, Physics and Informatics, Department of Nuclear Physics and Biophysics, 84248 Bratislava, Slovakia*
[b] *Institut für Kernphysik, Westfälische Wilhelms Universität, Münster, Germany*
[c] *Joint Institute for Nuclear Research, Dubna, Moscow Region, Russian Federation*

**Abstract** Evaluation of single, coincidence and anticoincidence gamma-ray spectrometry methods has been carried out with the aim to search for double-electron capture in $^{74}$Se. This process is unique, because there is probability for transition to the $2^+$ excited state in $^{74}$Ge (1204 keV), and de-excitation through two gamma-quanta cascade with energies of 595.9 keV and 608.4 keV. Long-term measurements with anticosmic shielded HPGe spectrometer and the coincidence HPGe-NaI(Tl) spectrometer did not show any evidence for the double-electron capture in $^{74}$Se. The best limit for the half-life of the double electron capture in $^{74}$Se (both for the neutrinoless and two neutrino processes) was estimated to be $>1.5 \cdot 10^{19}$ years.

**Keywords** HPGe spectrometer, Background of HPGe detectors, Double electron-capture decay, Selenium-74

## 1. Introduction

Observation of neutrino oscillations have provided evidences for nonzero neutrino mass and neutrino mixing, but these experiments are not sensitive to neutrino mass, so other processes are needed to experimentally estimate the neutrino mass. To reach this goal, precious studies of rare neutrinoless double beta ($\beta\beta(0\nu)$) decays are needed. Several large-scale underground experiments have been carried out [1-3], and several new ones are under construction [3-5].

Much less attention has been given, however, to the investigation of $2\beta^+$, $\beta^+$EC (EC = Electron Capture) and double EC (ECEC) processes, mainly because of the fact that predicted half-lives were outside of experimental possibilities. A review of experimental investigations of ECEC processes has been published by Barabash [6], who emphasized that recent developments in low-background experiments could make such studies feasible. This double β-decay process can take place either with emission of 2 neutrinos, ECEC(2ν), or without neutrinos ECEC(0ν). It is difficult to investigate the ECEC decay to the ground state (except of *in situ* experiments with Kr or Xe sources in the form of noble gases or liquids) because emitted low energy X-rays are absorbed in the sample. However, the ECEC process may be accompanied by transitions to excited states of daughter nuclei, which may be detected as a cascade of γ-quanta [7].

Recent theoretical predictions for the half-life of the ECEC(0ν) decay are in the range of $10^{22}$-$10^{32}$ years [8], and some of the decays may be close to experimental possibilities. It has been shown that a resonance condition can exist for transition to the excited level of the daughter nucleus, which may enhance the transition rate by a factor of ~$10^6$. The ECEC(0ν) process should be thus competitive with $\beta\beta(0\nu)$ decay, and should be therefore experimentally investigated. The existence of the ECEC(0ν) process would mean, similarly as for the $\beta\beta(0\nu)$ decay that lepton number is violated and the neutrino is a Majorana particle, what would require new particle physics beyond the Standard Model.

There are 12 nuclei which can undergo only the ECEC decay. From the experimental point of view the most interesting nuclei are those ($^{74}$Se, $^{78}$Kr, $^{96}$Ru, $^{106}$Cd, $^{112}$Sn, $^{130}$Ba, $^{136}$Ce and $^{162}$Er) for which resonance transition to the excited states of daughter nuclei is possible [6,9]. The first positive result obtained was a geochemical experiment with $^{130}$Ba where a half-life of the ECEC(2ν) decay process was estimated to be $(2.2\pm0.5) \, 10^{21}$ yr [10]. Recently a noble gas experiment using $^{78}$Kr as the source claimed a first positive result for the direct experiment determining the half-life of the ECEC(2ν) decay to be $(9.2^{+5.5}_{-2.6}$ (stat)±1.3(syst)) $10^{21}$ yr [11].

---

* Corresponding author. Comenius University, Department of Nuclear Physics and Biophysics, Mlynska dolina F-1, 84247 Bratislava, Slovakia.
Tel: +421 260295544; Fax: +421 265425882; Email address: povinec@fmph.uniba.sk

We shall focus in this paper on a search for double-electron capture in $^{74}$Se, where transitions to the excited states of daughter nuclei have been proposed, which are from the experimental point of view easier to detect than transitions to the ground states. Frekers [7] has shown that $^{74}$Se could be a good candidate for searching for the neutrinoless ECEC process because of a suitable decay scheme and energy. He described a neutrinoless double electron capture in $^{74}$Se through the degeneracy of the parent ground state with the second excited state in the daughter nucleus $^{74}$Ge. The mechanism of this process has already been described in detail [12] therefore we shall not repeat it here. We only notice that the decay energy of $^{74}$Se has recently been re-measured precisely to 1209.169 keV [13]. The decay energy coincides within 5 keV with the second excited state of the $^{74}$Ge daughter at 1204.2 keV. This state de-excites with about 70% probability through a 608.4 keV and 595.9 keV γ-ray cascade to the ground state of $^{74}$Ge.

High detection efficiency of the present day HPGe detectors (up to 200% relative efficiency for the $^{60}$Co 1332.5 keV γ-rays measured by the 7.6 cm diameter and 7.6 cm long NaI(Tl) detector), and their excellent energy resolution permits selective and non-destructive analyses of several radionuclides in samples of various origin [14-17]. As the efficiency and resolution of these detectors is limited, usually only the way of improving their sensitivity is in decreasing their background. The background is mainly composed of cosmic radiation (hard and soft components) and radiation coming from radionuclides found around and inside the detector. In underground laboratories both the soft (mostly electrons and positrons) and hard (mostly muons) components of the background can be significantly reduced from the whole cosmic-radiation background [18,19]. However, in a surface laboratory the reduction of the hard component of cosmic radiation is very limited because thick layers of matter are required to shield a detector [20-24]. Another possibility for a partial reduction of the detector background from muons in surface and shallow underground laboratories is to use coincidence/anticoincidence arrangements of the detectors [24,25]. A coincidence γ-ray spectrometer may be a choice when radionuclides emitting positrons (e.g. $^{22}$Na, $^{26}$Al) or cascade γ-quanta (e.g. $^{60}$Co) should be analyzed [14-17,26,27].

The low-level γ-ray spectrometry has been applied in searches for double beta-decay or double electron capture of several nuclides [12,16,28]. The first measurements of selenium as a candidate for double-electron capture decay were carried out in the Modane (France) underground laboratory (operating at the depth of 4800 m water equivalent), using a HPGe detector (400 cm$^3$, resolution 2.0 keV @ 1.33 MeV). The best limit on the half-life of the double electron capture in $^{74}$Se (using 563 g of selenium powder, containing of 4.69 g of $^{74}$Se, analyzed during 437 hours) was $> 0.55 \cdot 10^{19}$ years [16]. First our measurement using a HPGe-NaI(Tl) coincidence γ-ray spectrometer (operating in a surface laboratory in Bratislava) gave a limit on the half-life of $> 0.14 \cdot 10^{19}$ years [12].

The aim of the present work has been to evaluate single – coincidence – anticoincidence modes of operation of γ ray spectrometers, and to analyze the same selenium sample as in the previous experiment [11] using a HPGe γ-spectrometer operating in a surface laboratory with and without an anticosmic shielding, as well as using a HPGe-NaI(Tl) coincidence γ-ray spectrometer for a longer counting time as in the previous experiment.

## 2. Materials and methods

*2.1. HPGe γ-spectrometer with passive and anticosmic shieldings*

As already mentioned our goal has been to decrease the background of the HPGe spectrometer to a minimum using a heavy passive shielding and an active anticosmic shielding. To keep radioactive contamination of the detector material at minimum is not a problem for HPGe detectors as they require high-purity Ge crystals for proper operation. However, the cryostat and associated materials around the HPGe detector, as well as the passive shielding against cosmic rays and radiation from surrounding objects (e.g. walls of the laboratory, glass windows, etc.) can be contaminated by natural radionuclides. The Central Europe is still contaminated by anthropogenic radionuclides from the Chernobyl fallout, mainly by $^{137}$Cs [29]. The background of the HPGe γ-spectrometer operating in a shield with large inner volume can also be affected by temporal variations of natural radionuclides in the air, mainly due to radon ($^{222}$Rn) and thoron ($^{220}$Rn) decay products, therefore, special arrangements are necessary to minimize and stabilize the radon content in the shield [17].

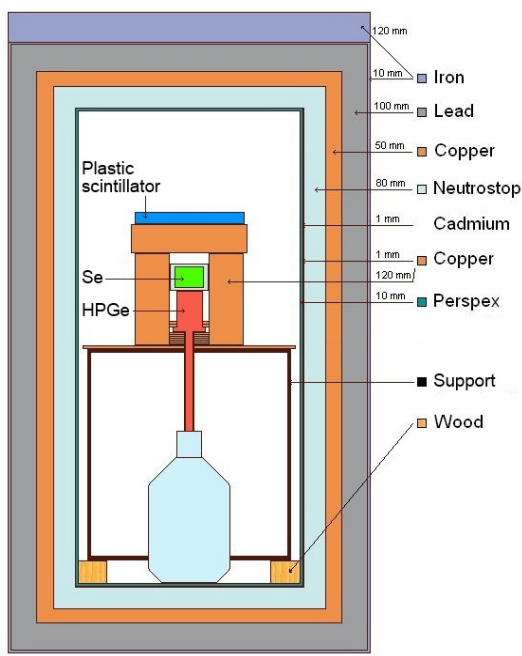

**Fig. 1** Low-level HPGe detector with passive (iron+lead+copper) and active (anticosmic) shielding with plastic scintillation detector

Large volume shield designed for low-level gamma-ray spectrometry, with our dimensions of 1.5 x 1.5 x 2 $m^3$, was used during measurements (Fig. 1). The vertical dipstick of the HPGe detector has been used in the shield design with the aim to allow for more versatile two parameter coincidence/anticoincidence γ-spectrometer with large cylindrical plastic scintillation detectors, also used as an anticompton shielding [30].

The shield consists of the following layers (from the outside to the inside): 10 cm of lead, 10 cm of electrolytic copper, 10 cm of polyethylene with boric acid, 0.1 cm of electrolytic copper, 0.1 cm of cadmium and 1 cm of perspex. On the top of the shield, a layer of 12 cm of iron is added. The inner dimensions of the shield are 0.80 x 0.90 x 1.72 $m^3$. The polyethylene with boric acid is slowing down cosmic-ray neutrons (as well as neutrons originating in the iron and copper shield by interactions of muons), which may be then captured by boron nuclei. The layer of polyethylene with boric acid decreased the neutron flux in the shield by about 30%.

In order to further decrease the background of the HPGe spectrometer, an extra internal copper shield (12 x 20 x 30 $cm^3$) has been inserted inside the large shield (Fig. 1). Contributions of the preamplifier and the Dewar (filled with liquid nitrogen) to the HPGe γ-spectrometer background have been decreased by an additional copper shield inserted at the bottom of the cryostat. All materials used inside the shielded chamber have been carefully selected with minimum radionuclide contamination. The copper sheets and blocks were made from electrolytic copper with low radioactive contamination. Measured activities of the main primordial radionuclides in the copper were low ($^{40}$K: <40; $^{235}$U: <5; $^{238}$U: <35; $^{232}$Th: <10; all results in mBq $kg^{-1}$). Activity of the cosmogenic $^{60}$Co produced in the copper shield by cosmic rays was <3 mBq/kg.

A HPGe detector produced by PGT Europe (a sensitive volume of 270 $cm^3$) with 70% relative efficiency for the $^{60}$Co 1332.5 keV gamma-rays (relative to 7.6 cm diameter and 7.6 cm long NaI(Tl) detector) operated in this shield. Its energy resolution was 1.9 keV (FWHM at 1332.5 keV). A detail description of the shield characteristics has already been published [16,23] therefore it will not be described here.

During coincidence measurements an NaI(Tl) scintillation detector (10 cm in diameter and 10 cm long) was placed above the selenium sample (in this case the top copper plug was removed from the shield. The resolution of the NaI(Tl) detector was 6.4% at 661.6 keV.

To further decrease the background of the single HPGe γ-ray spectrometer, an anticosmic shielding made of plastic scintillation detector (40 x 40 x 5 $cm^3$) has been installed on the top of the inner copper shield (Fig. 1). The aim of the anticosmic shielding has been to decrease a contribution of cosmic-ray muons to the HPGe γ-spectrometer background, as predicted by Monte Carlo simulations [24,25,31]. The low background HPGe coincidence-anticoincidence γ-spectrometer is located in the basement of the Faculty of Mathematics, Physics and Informatics of the Comenius University in Bratislava. The concrete thickness above the low-level shield has been about 1.2 m. The walls of the laboratory were made of concrete with 0.3 m thickness, and the bottom iron-concrete layer was about 0.4 m thick.

A schematic diagram of the electronics used throughout the measurements, composed of CANBERRA NIM modules is presented in Fig. 2. A typical stability of the electronics was ± 1 channel with range up to 4000 channels. A daily check on the apparatus assured that the performance of the electronics was within the expected range of variations.

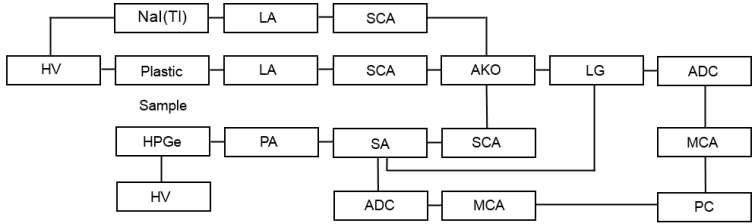

**Fig. 2** Schematic view of the coincidence-anticoincidence electronics used in the experiments (CANBERRA NIM modules). HV – high voltage power supply, LA – linear amplifier, PA – preamplifier, SA – spectrometric amplifier, SCA – single channel analyzer, AKO – anticoincidence circuit, LG – linear gate, ADC – analog to digital converter, MCA – multichannel analyzer, PC – computer.

A sample from natural selenium (purity of 99.99%) was used to study double-electron capture decay of $^{74}$Se. The abundance of $^{74}$Se in the natural selenium is 0.89%, what corresponds to 27 g (or $2.2 \cdot 10^{23}$ atoms) of $^{74}$Se in the 3 kg sample. The sample density was 1.4 g/cm$^3$. Selenium sample was in the form of grains with diameter of 2 mm, which were placed in a plastic container with the inner diameter of 12 cm and the inner height of 8 cm (16 cm the outer diameter and 9.5 cm the outer height). The selenium sample was before the analysis cleaned in an ultrasonic bath with ethylalcohol, dried and kept under nitrogen atmosphere in the sample container.

*2.2. Calibration*

A multi-radionuclide etalon source has been used for the manual energy calibration. The efficiency calibration was carried out using Monte Carlo simulations with GEANT 3.21 software package [32], and also with etalons of known activity placed inside the same geometry as the selenium sample. Estimation of the dead layer of the HPGe detector is important input for Monte Carlo simulations of the detector efficiency. The detector dead layer was estimated using a set of point radioactive standards in the wide range of γ-ray energies from 60 to 1808 keV in different positions and distances from the detector window. The measurements were performed with the collimated and uncollimated γ-rays. The estimated thickness of the dead layer (1.2 mm) was similar to other p-type HPGe detectors. In the Monte Carlo simulations the dead layer was treated together with the thickness of the detector endcap (1.4 mm of aluminum), and the free space between the Ge crystal and the endcap (3 mm). Fitting the experimental efficiency points with an efficiency curve, and comparing it with Monte Carlo simulated efficiency curve we found an agreement within 7%.

The calculated Monte Carlo efficiencies (Table 1) are from 0.88% at 595.8 keV to 0.63% at 1204.2 keV for the single HPGe spectrometer, and from 0.009% to 0,003% for the coincidence HPGe-NaIT(Tl) spectrometer (for the same energy range). As in this experiment the selenium sample was of a regular cylindrical shape, systematic uncertainties were smaller than in the case of irregular sample shapes [33,34]. The total uncertainty of the Monte Carlo simulation efficiency method used in half-life calculations, comprising of statistical (4%) and systematic uncertainties (5%), was 7%. A detail description of the Monte Carlo and experimental efficiency calibration methods of γ-ray spectrometers for non-destructive analysis of large volume samples has already been published [33,34].

**Table 1**
Detection efficiencies of the single HPGe and coincidence HPGe-NaI(Tl) gamma-spectrometers used for the measurements of $^{74}$Se gamma-rays.

| Energy (keV) | ε (Single HPGe spectrometer) | ε (Coincidence HPGe-NaIT(Tl) spectrometer) |
|---|---|---|
| 595.8 | 0.00884 | $9.1 \times 10^{-5}$ |
| 608.4 | 0.00873 | $9 \times 10^{-5}$ |
| 1204.2 | 0.00626 | $3.1 \times 10^{-5}$ |

## 3. Results and discussion

*3.1. Background studies of the HPGe γ-spectrometer*

Three measuring modes of the low-level γ-ray spectrometry used for the analysis of the selenium sample were investigated: a single HPGe γ-spectrometer a HPGe γ-spectrometer with anticosmic shielding, and a coincidence HPGe-NaI(Tl) γ-spectrometer. As discussed in [12,16], the dominant transition in double electron capture of $^{74}$Se should be a de-excitation of the second excited state of the $^{74}$Ge daughter at 1204.2 keV via a γ-ray cascade to the ground state of $^{74}$Ge with emissions of 608.4 keV and 595.9 keV γ–rays. The main γ–ray peaks, which should be searched for in experimental γ–ray spectra, are therefore 608.4 keV, 595.9 keV and at 1204.5 keV. Unfortunately, as the energy resolution of the HPGe detector is 1.9 keV (FWHM at 1332.5 keV), the peak of interest at 608.4 keV cannot be distinguished from the background peak at 609.3 keV, which is due to the decay of the $^{214}$Po, populated by the $^{214}$Bi β−decay, either directly or through a rather complicated γ-ray cascade.

The main sources of background in this experiment are decay products of $^{222}$Rn (mainly $^{214}$Bi) and $^{220}$Rn (mainly $^{208}$Tl). A special effort was made therefore to minimize the free space near the detector using the inner copper shielding. A further reduction in the background and its stabilization during long-term measurements (due to possible radon variations in the laboratory) was accomplished by flushing the copper shield volume with nitrogen evaporating from the detector Dewar [17]. The background due to $^{214}$Bi contamination of the experimental set up is crucial for the experiment. There are two sources of $^{214}$Bi γ-quanta observed in the background spectrum:

(i) Radon emanating in the low-level shield from construction materials, e.g. from copper walls, cryostat, preamplifier, selenium sample and its container, etc. Although the HPGe detector with the selenium sample and the inner copper shield was hermetically closed by a polyethylene foil and continuously flushed by nitrogen gas evaporating from the Dewar, it was not possible to remove completely radon from the space surrounding the HPGe detector.

(ii) $^{214}$Bi γ-rays originating directly from the above mentioned construction materials, and irradiating the HPGe detector.

The measured background spectra in the energy region of interest have been compared for single-coincidence-anticoincidence modes with the aim to study the background characteristics of the HPGe γ-spectrometer. The simplest geometry used in the experiments was the single HPGe γ-spectrometer. Figure 3 compares the measured γ-ray spectra of the selenium sample and of the background (i.e. without the selenium sample). The chosen energy range was 550-700 keV and 1100-1300 keV. The main background peaks from $^{214}$Bi are clearly visible in both energy regions (609.3 and 1238 keV, respectively).

The γ-ray spectra measured with the selenium sample are more suppressed due to additional shielding of the HPGe detector by the sample itself. The large selenium sample represents another solid absorber with high atomic number, that absorbs part of the background associated with cosmic-ray secondary particles, as well as a radiation from materials located above the HPGe detector. Peaks, representing contributions from thoron (583 keV) and radon (609.3 keV and 1238 keV) decay products to the HPGe γ-spectrometer background are also suppressed by the large selenium sample (Fig. 3).

Contributions from cosmic rays will affect all peaks of interest, while the radon contribution will be mainly via the Compton continuum from its daughter $^{214}$Bi. The total background in the energy range 400-1400 keV was measured to be 18980 ± 90 counts day$^{-1}$ kg$^{-1}$ of Ge (the counting time was 112 days). Such a low background was achieved because of internal shielding made of electrolytic copper with low radioactive contamination. The obtained background is in agreement with Monte Carlo simulations published in separate papers [24,25].

To preserve a high detection efficiency of the system, and to additionally suppress the HPGe detector background, an anticosmic veto shielding made from plastic scintillation detector has been applied. A comparison of γ-ray spectra of the selenium sample measured with HPGe detector with and without anticosmic veto shielding (Fig. 4) shows that 37% background suppression was achieved in the energy region from 550 to 650 keV.

The total background in the anticoincidence mode calculated for the energy region 400-1400 keV was 11050 ± 60 counts day$^{-1}$ kg$^{-1}$ of Ge, what represents 58% background suppression. Table 2 compares background counting rates under different peaks for single and anticoincidence modes of measurements. The best factor of merit ($F = \varepsilon/\sqrt{b}$) have been obtained for the anticoincidence mode for the peaks of 595.8 and 1204.2 keV (0.07 and 0.09, respectively), better than for the single spectrometer (0.05 and 0.07, respectively).

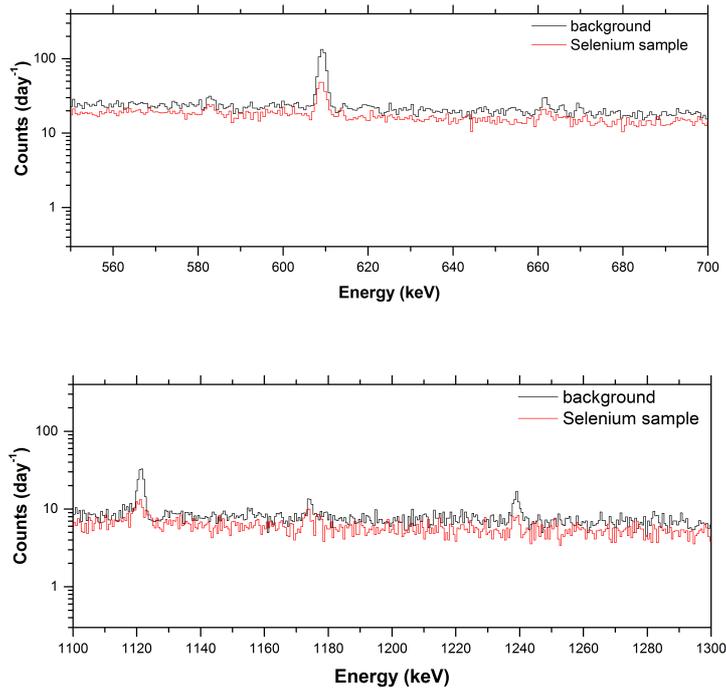

**Fig. 3** Gamma-ray spectra of background (top) and the selenium sample (bottom) by the single HPGe detector in the region 550-700 keV and 1100-1300 keV (measuring time: 8 days for background and 7 days for the sample).

The background peaks due to radon (609.3 keV from $^{214}$Bi) and thoron (583 keV from $^{208}$Tl) decay products have also been suppressed when compared with peaks measured with the single HPGe spectrometer (Fig. 3), but they are still visible. In the higher energy region, the peaks with the energy of 1120 keV and 1238 keV are both from the $^{214}$Bi decay as well. It is interesting to note that we also observed a peak with energy of 1173 keV (from $^{60}$Co), which is a cosmogenic product of the interaction of cosmic-ray nucleons with the shielding material. A peak with energy of 661.6 keV is due to $^{137}$Cs of the Chernobyl origin, which is still present in our region [17,29].

**Table 2**
Comparison of background under peaks of interest for investigation of double electron capture in $^{74}$Se.

| Counts (day$^{-1}$) | 590.1 keV | 595.8 keV | 601.4 keV | 608.4 keV | 611.1 keV | 1204.2 keV |
|---|---|---|---|---|---|---|
| Single HPGe spectrometer | 250 | 260 | 260 | 390 | 390 | 79 |
| HPGe spectrometer with anticosmic shielding | 170 | 175 | 180 | 270 | 250 | 51 |

The lowest background was achieved with the HPGe–NaI(Tl) coincidence γ-spectrometer (Fig. 5). With the coincidence spectra measured for 105 days the total background calculated for the energy region 400-1400 keV was 345 ± 15 counts day$^{-1}$ kg$^{-1}$ of Ge. As it can be seen from Fig. 5, the greatest advantage of the coincidence mode is the suppression of background peaks from $^{214}$Bi and $^{208}$Tl.

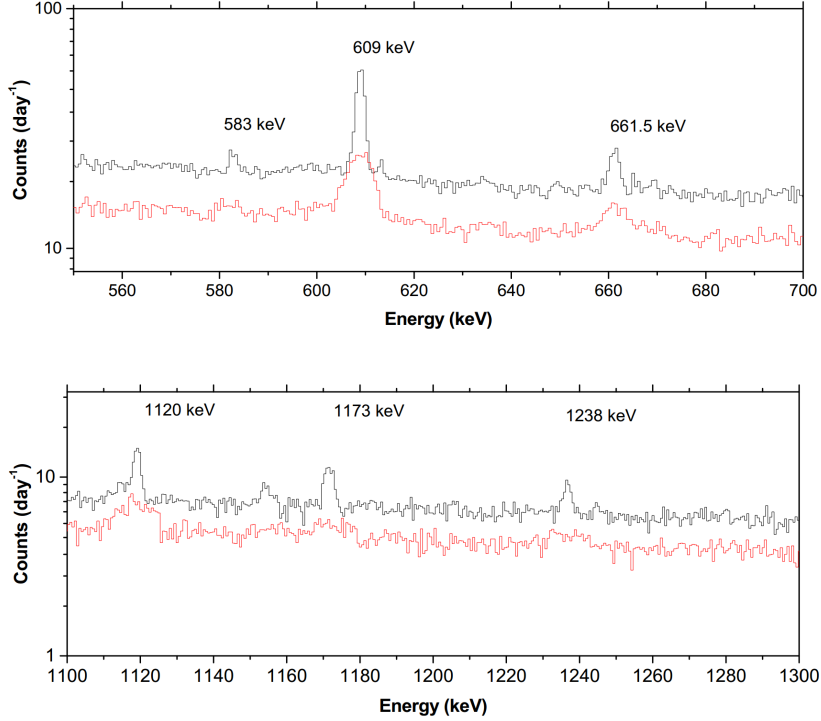

**Fig. 4** Comparison of single HPGe (top) and anticoincidence (bottom) gamma-ray spectra of the selenium sample in the region 550 -700 keV and 1100-1300 keV (measuring time: 34 days for both spectra)

It has been found, however, that for large-volume samples (as it is in this case) the coincidence geometry suffers in detection efficiency because of the large distance between the HPGe and NaI(Tl) detectors, as well as due to the self-absorption of investigated γ-rays in the sample. Therefore the factor of merit for the coincidence mode is much lower than for the anticoincidence mode (F = 0.002 for the 595.8 - 608.4 keV coincidences).

*3.2. Double-electron capture in $^{74}$Se*

Several transition processes in the double electron capture decay of $^{74}$Se are possible [16], however, we shall focus only on those, which are well defined and have highest probabilities to be measured. For the neutrinoless double electron-capture decay of $^{74}$Se it is the process of the capture of two electrons from the L shell with transition to the $2^+$ level of $^{74}$Ge (1204.2 keV), accompanied by 595.8 keV and 608.4 keV de-excitation γ- quanta (66%), or a 1204.2 keV de-excitation γ–quantum (34%). For the two neutrino double electron-capture decay of $^{74}$Se, accompanied by γ-quanta, it is the transition to the $2^+$ level of $^{74}$Ge with emission of γ-quanta of 595.8 keV, 608.4 keV (66%) and 1204.2 keV (34%). We shall evaluate half-lives of these decays separately for the single/anticoincidence mode and for the coincidence mode of measurements.

*3.2.1. Single and anticoincidence γ-ray spectrometry*

Gamma-ray spectra of the selenium sample measured with HPGe detector with and without anticosmic veto shielding used for searching for ECEC(0ν) decays are presented in Fig. 4. The data evaluation is complicated by the fact that the γ-ray spectra measured with the selenium sample have lower counting rates (a lower background) in the full energy range due to additional shielding of the HPGe detector with the selenium sample. For estimation of the background in the energy region of interest we used a traditional least square fit after rejecting all well visible peaks in the spectrum (i.e. the annihilation peak at 511 keV, and the peaks from $^{208}$Tl and $^{214}$Bi present in the spectrum due to radioactive contamination of the detection system). In the case of the transition peak at 608.4 keV, which cannot be separated with the present HPGe detector resolution from the $^{214}$Bi background peak at 609.3 keV, the counting rate under this peak has been used for estimation of the half-life. The best background conditions are for the resonant transition to the 1204.2-keV excited state of $^{74}$Ge (Fig. 4).

For the single HPGe spectrum (the energy resolution of 1.9 keV), 8 channels in the spectrum were considered for the calculation of the half-life limit. For the anticoincidence spectrum (the resolution worsened to 2.9 keV), 12 channels in the spectrum were used for the calculations. Background counting rates under the peaks of interest used for calculations of the half-life limits are compared in Table 2. The total uncertainty of the results was, as it is typical in low-level experiments, mainly associated with counting statistics (below 15%). The systematic uncertainty was mainly due to Monte Carlo efficiency calibration (7%), and possible discrepancies in nuclear data (3%). The total uncertainty of the results was estimated to be below 17%.

It can be seen From Fig. 4 that there are no statistically significant peaks at the energies of interest. Therefore we can estimate only half-life limits calculated from the background counting rates under the expected peaks. Limits have been calculated for different γ-lines corresponding to the transitions under study.

The half-life limit of the ECEC(0ν) decay process was calculated using the formula [28]

$$T > \alpha R \varepsilon \{(m\,t)/(b\,\delta E)\}^{1/2}$$

where $\alpha$ is the constant (including ln 2, Avogadro number, the isotopic abundance of the $^{74}$Se and its molar mass; in our case $\alpha = 8.8 \times 10^{21}$), $R$ is the branching ratio of the nuclear transition under study, $\varepsilon$ is the detection efficiency, $m$ is the mass of the selenium sample, $t$ is the time of measurement, $b$ is the background counting rate measured with the selenium sample under different peaks, and $\delta E$ is the energy window of the ECEC(0ν)-decay events (proportional to the energy resolution of the detector).

The calculated half-life limits of the double electron capture in $^{74}$Se are presented in Table 3. The longest half-life limits (both for the neutrinoless and two neutrino electron capure), derived with 90% confidence, have been obtained with anticoincidence ($T_{1/2} > 1.5\ 10^{19}$ yr) and single ($T_{1/2} > 0.8\ 10^{19}$ yr) γ-spectrometers for the process with the emission of 595.8 keV, 608.4 keV and 1204.2 keV de-excitation γ –quanta.

### 3.2.2. Coincidence γ-ray spectrometry

Several transition possibilities of the ECEC(0ν) decay of $^{74}$Se to excited states were considered [28,30]. We shall focus in this work on double capture of L-electrons with transitions to the $2^+$ level of $^{74}$Ge (1204.2 keV), accompanied by 595.8 keV and 608.4 keV de-excitation γ-quanta, or a 1204.2 keV de-excitation γ-quantum.

**Table 3**
Half-lives lower limits for double-electron capture in $^{74}$Se.

| Transition | Energy of γ-rays (keV) | HPGe-NaI(Tl) coincidence spectrometer ($10^{19}$ yr) | Single HPGe spectrometer ($10^{19}$ yr) | HPGe spectrometer with anticosmic shielding ($10^{19}$ yr) | Previous works ($10^{19}$ yr) |
|---|---|---|---|---|---|
| ECEC(0ν) | 595.8 608.4 | 0.4 | | | 0.14* [10] |
| ECEC(0ν) | 595.8 608.4 1204.2 | | 0.8 | 1.5 | 0.55 [14] |
| ECEC(2ν) | 595.8 608.4 1204.2 | | 0.8 | 1.5 | 0.55 [14] |
| Final ECEC(0ν) | | | | 1.5 | |
| Final ECEC(2ν) | | | | 1.5 | |

*) corrected for the coincidence γ-spectrometer efficiency

The full coincidence γ-ray spectrum of the selenium sample used for calculation of the half-life of the ECEC(0ν) decay of $^{74}$Se (coincidences between 595.8 keV and 608.4 keV γ-quanta) measured during 105 days with the HPGe detector in coincidence with the NaI(Tl) detector, is presented in Fig. 5. Unfortunately, the search for the 595.8 keV and 608.4 keV γ-ray cascade in double electron capture in $^{74}$Se has not been succesful yet. There are no statistically significant peaks at these energies, therefore only half-life limits of the ECEC(0v) decays could be estimated. The calculated half-life limit, derived with 90% confidence, is above 0.4 10$^{19}$ yr (Table 3), about a factor of 3 higher than in our previous work [12].

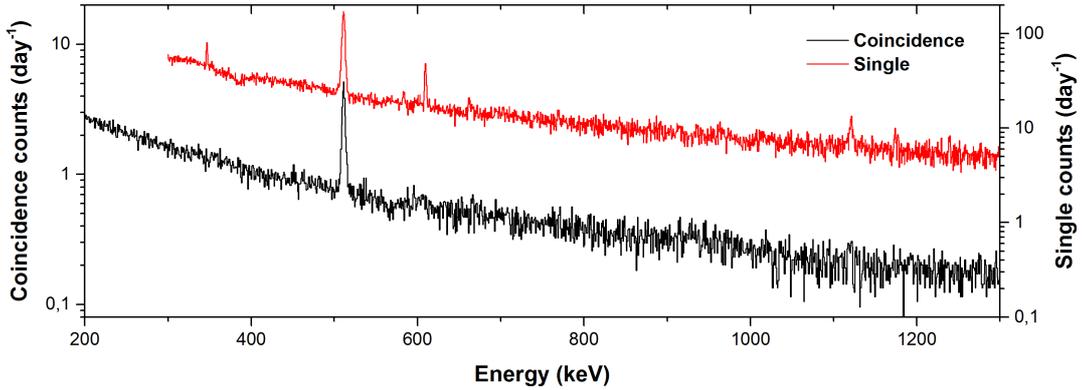

**Fig. 5** Coincidence gamma-ray spectrum (coincidences between 595.8 keV and 608.4 keV gamma-quanta) measured with the HPGe-NaI(Tl) coincidence spectrometer (measuring time: 105 days), and a single HPGe spectrometer gamma-ray spectrum (top) of the selenium sample (measuring time: 34 days).

*3.2.2. Discussion on half-life limits*

The best value for the limit of half-lives on the double electron capture process in $^{74}$Se (both with and without neutrinos) >1.5 10$^{19}$, derived with 90% confidence (Table 3), is by about factor of 3 higher than in the previous measurement obtained with a smaller sample and during shorter measuring time in the Modane underground laboratory [16], and by about a factor of 10 higher than in our previous experiment [12].

Recently the experimental search for a resonance transition in $^{112}$Sn was investigated [9] and a limit of $T_{1/2}$ > 4.7 10$^{20}$ yr (90% confidence) was obtained for the neutrinoless double electron-capture transition to the $0^+$ state at 1871 keV. Similar experiments carried out recently with $^{96}$Ru and $^{106}$Cd sources gave the limits >1.3 10$^{19}$ yr [35] and >1.6 10$^{20}$ yr [36], respectively. Very recent liquid xenon experiment gave for the ECEC(2v) decay of $^{124}$Xe to the ground state a half-life limit >1.66 10$^{21}$ years at a 90% confidence level [37].

As the double electron-capture effect is a very weak process, and mostly only limits on half-lives could be estimated till now, the search for this process should continue. This is specifically important in the context of possible enhancement of the decay rate by several orders of magnitude as the neutrinoless double electron-capture process is nearly degenerate with an excited state in the daughter nuclei. The detection of the neutrinoless double electron capture is especially important as it could help to determine the effective Majorana neutrino mass, and parameters of right-hand current admixture in electroweak interaction. Detection of the two neutrino double electron capture would be also important as it would help to determine the nuclear matrix elements, important for calculations of half-lives of other double β-decay processes.

The presented results clearly indicate that the γ-spectrometer background is playing a crucial role in these experiments. A better limit for this process should be possible to achieve by operating a low-level HPGe spectrometer (or a two parameter HPGe-HPGe coincidence spectrometer) in a deep underground laboratory during at least one year of counting time, and using a selenium sample enriched in $^{74}$Se (or other samples with enriched nuclei, e.g. $^{96}$Ru, $^{106}$Cd, $^{112}$Sn, $^{130}$Ba, $^{136}$Ce and $^{162}$Er) for which resonance transition to the excited states of daughter nuclei is possible [6].

**4. Conclusions**

The main results presented in this work may be summarized as follows:
(i) Low-level HPGe γ-spectrometer operating in single, anticoincidence and coincidence modes for analysis of the large-volume sample of natural selenium has been described with the aim to search for double electron capture in $^{74}$Se. The best factor of merit has been reached with the anticoincidence γ-spectrometer operating with plastic scintillation detector as an anticosmic shielding, which decreased the detector background by about 60%, and kept its high detection efficiency. The great advantage of the coincidence mode with HPGe and NaI(Tl) detectors has been the suppression of background peaks from the radon ($^{214}$Bi) and thoron ($^{208}$Tl) daughters. It has been found, however, that for large-volume samples the coincidence geometry suffers in detection efficiency because of the large distance between the detectors, as well as due to the self-absorption of investigated γ-rays in the sample.

(ii) The longest half-life limits, both for the neutrinoless and two neutrino electron capure in $^{74}$Se, derived with 90% confidence, have been obtained with antocoincidence ($T_{1/2} > 1.5 \ 10^{19}$ yr) and single ($T_{1/2} > 0.8 \ 10^{19}$ yr) γ-spectrometers for the process with the emission of 595.8 keV, 608.4 keV and 1204.2 keV de-excitation γ–quanta. This result is by about factor of 3 higher than previous measurement obtained with a smaller sample and during shorter measuring time in the Modane underground laboratory [16], and by about a factor of 10 higher than in our previous experiment [12].

A better limit for a double electron-capture process should be possible to achieve in the future experiments operating a low-level HPGe (or two parameter HPGe-HPGe coincidence) spectrometer in a deep underground laboratory (e.g. at Modane or Gran Sasso) during at least one year counting time, and using a selenium sample enriched in $^{74}$Se, or other samples with enriched nuclei for which resonance transition to the excited states of daughter nuclei is possible. Such experiemnnts are of great importance as they could help to solve the problem if neutrino is a Majorana or Dirac particle (in the case of successful neutrinoless double electron-capture experiment), or they would help to determine the nuclear matrix elements, important for calculations of half-lives of other double β-decay processes (in the case of successful two neutrino double electron-capture experiment).

## Acknowledgements

The Bratislava group acknowledges a partial support provided by the VEGA grant No. 1/0783/14 from The Ministry of Education, Science, Research and Sport of the Slovak Republic, as well as by the EU Research and Development Operational Program funded by the ERDF (projects Nos. 26240120012, 26240120026 and 26240220004).